\documentclass[nofootinbib,twocolumn,preprintnumbers,amsmath,amssymb]{revtex4}
\pdfoutput=1 

\begin{document}

\title{\boldmath Gravitomagnetism in  MOdified theory of Gravity }


\author{Qasem \surname{Exirifard}}


\affiliation{ The Abdus Salam International Centre for Theoretical Physics (ICTP), Strada Costiera, Trieste, Italy
}

\email{exirifard@gmail.com}

\begin{abstract}
We study the gravitomagnetism in the Scalar-Vector-Tensor theory or Moffat's Modified theory of Gravity(MOG). We compute the gravitomagnetic field that a slow-moving mass distribution produces in its Newtonian regime. We report that the consistency between the MOG gravitomagnetic field and that predicted by the Einstein's gravitional theory and measured by Gravity Probe B, LAGEOS and LAGEOS 2, and with a number of GRACE and Laser Lunar ranging measurements requires $|\alpha| < 0.0013$.  We provide a discussion.
\end{abstract}

\maketitle

 The human civilisation still lacks a conclusive knowledge about the physics at galaxies and beyond, in the sense that either dark matter exists or the dynamics of gravity is more involved than that predicted by the Einstein's gravitional theory. In the first approach, we have not yet found any dark matter. The later approach started by the Milgrom's theory of the Modified Newtonian Dynamics (MOND)\cite{MOND}, changed to the a-quadratic Lagrangian model for  gravity  (AQUAL)\cite{AQUAL}.  A generally covariant realisation of the AQUAL model is the TeVeS theory  \cite{Bekenstein:2004ne}.  TeVeS is in agreement with the Gravitomagnetism and the near horizon geometry of super massive black hole \cite{Exirifard:2011vb}. It has been shown that changing EH gravity to TeVeS can not reproduce the CMB \cite{Skordis:2005xk}. Though adding a $4^{\text{th}}$ sterile neutrino for $11-15~eV$ to matter content of TeVeS theory fits the spectrum in a LCDM cosmology \cite{Angus:2008qz}, it wouldn't form the large structure of the universe \cite{Xu:2014doa}. TeVeS also appears not to be consistent with GW170817 \cite{Boran:2017rdn}. There is also Moffat's theory of gravity  \cite{Moffat:2005si} that passes many tests and observations when TeVeS fails; particularly it produces the acoustic peaks in the cosmic microwave background radiation and the matter power spectrum \cite{Moffat:2007ju}, and it is consistent with gravitational wave's data \cite{Green:2017qcv, Rahvar:2018nhx}. 
 We would like to study the gravitomagnetism in Moffat's theory.   In the first section we review Moffat's theory of gravity  around the Earth space-time geometry\cite{Moffat:2014aja}. In the second section we present the gravitomagnetism approximation to Moffat's theory around the space-time of the Earth. We report how Gravity Probe B \cite{Everitt: }, LAGEOS and LAGEOS 2\cite{LAGEOS}, and with a number of GRACE and Laser Lunar ranging measurements \cite{Murphy:2007nt} constrain the  $\alpha$ parameter to $|\alpha|<0.0013$.  In the last section we provide a discussion.

\section{The action of MOG theory around the Earth}
The action of MOG theory \cite{Moffat:2005si}  is given by:
\begin{eqnarray}
S = \int d^4x (L_G + L_\Phi + L_S + L_M)\,,
\end{eqnarray} 
where $L_M$ is the Lagrangian density of the matter and
\begin{eqnarray}
L_S &=& \sqrt{-g} 
\left\{
\frac{1}{G^3} [\frac{1}{2} g^{\mu\nu} \nabla_\mu G \nabla_\nu G - V(G)]\right.\nonumber
\\
& &~~~~\left.+ \mu^2 G [\frac{1}{2}g^{\mu\nu} \nabla_\mu \mu \nabla_\nu \mu - V(\mu)]
\right\} \,,\\
L_\Phi &=& \frac{1}{4\pi} \sqrt{-g} 
\left[ \frac{1}{4} B^{\mu\nu} B_{\mu\nu} -\frac{1}{2} \mu^2 \Phi^\mu \Phi_\mu + V({\Phi})
\right]\,,\\
L_G &=& \frac{1}{16\pi G} \sqrt{-g} (R+2\Lambda)\,, 
\end{eqnarray}
are the Lagrangian densities of the scalar, the vector and the tensor, respectively. $G(x)$ is related to the Newton's constant. $\mu(x)$ represents the mass of the vector field, $B_{\mu\nu}= \partial_\mu \Phi_\nu - \partial_\nu \Phi_\mu$, and $V(G), V(\mu), V(\Phi)$ are potential.

Following \cite{Moffat:2014aja}, ``we neglect the mass of the $\Phi_\mu$ field, for in the determination of galaxy rotation curves and galactic cluster dynamics $\mu=0.042\,(\rm kpc)^{-1}$, which corresponds to the vector field $\Phi_\mu$ mass $m_\Phi=2.6\times 10^{-28}$ eV~\cite{MoffatRahvar1,TothMoffat,MoffatRahvar2}. The smallness of the $\Phi_\mu$ field mass in the present universe justifies our ignoring it when solving the field equations for compact objects such as the vicinity of Earth,  neutron stars and black holes. However, the scale $\mu=0.042\,{\rm kpc}^{-1}$ does play an important role in fitting galaxy rotation curves and cluster dynamics".  Ref. \cite{Green:2019cqm}  provides the latest most detailed fitting of MOG to galaxy dynamics. Around the Earth space-time geometry $G_N$ is constant. So we set $\partial_\mu G=0$. Since $\mu$ is very small we ignore $\partial_\mu \mu$  at the vicinity of the Earth. We ignore the cosmological constant as well. Since the vector field around the space-time geometry of the Earth is small, and assuming that $V(\Phi)$ has a Taylor expansion around $\Phi_\mu=0$, we can ignore $V(\Phi)$ as well. These simplify Moffat's action  at the vicinity of the Earth to \cite{Moffat:2014aja}:
\begin{equation}\label{MoGLB}
S \,=\, \frac{1}{16\pi G} \int d^4 x \sqrt{-g} (R  + G B^{\mu \nu} B_{\mu\nu} )\,.
\end{equation}
The matter current density is defined in terms of the matter action $S_M$:
\begin{equation}
\frac{1}{\sqrt{-g}} \frac{\delta S_M}{\delta \Phi_\mu} \,=\, - J^\mu\,,
\end{equation}
where $J^\mu$ is proportional to the mass density. A test particle action is given by 
\begin{equation}\label{STP}
S_{TP}  = - m \int d\tau\sqrt{g_{\mu\nu} \dot{x}^\mu \dot{x}^\nu}  - k m \int d\tau \Phi_\mu \dot{x}^\mu \,,
\end{equation}
where dot is variation with respect to $\tau$ and $m$  denotes the test particle's mass. This means that geodesics in the Randers's geometry describes orbits of particles in the MOG \cite{Randers}.  The coupling constant is assumed to be proportional to the mass of the test particle and the proportionality factor reads:
\begin{subequations}
\label{alphak}
\begin{eqnarray}
k &=& \pm \sqrt{\alpha G_N}\,,\\
\alpha & =& \frac{G -G_N }{G_N}\,,
\end{eqnarray}
\end{subequations}
where $G_N$ is the gravitational Newton's constant.

Let an auxiliary field $e(\tau)$ be defined on the world line of the test particle. Then consider the following action:
\begin{equation}\label{STPxe}
S_{TP}[x,e] = - m \int d\tau (\frac{1}{2 e} g_{\mu\nu} \dot{x}^\mu \dot{x}^\nu + \frac{e}{2}) - k m \int d\tau  \Phi_\mu \dot{x}^\mu\,,
\end{equation}
the variation of which with respect to $e(\tau)$ yields:
\begin{equation}
\frac{\delta S_{TP}[x,e] }{\delta e} = 0 ~\to~ e(\tau) = \sqrt{g_{\mu\nu} \dot{x}^\mu \dot{x}^\nu}
\end{equation}
inserting which in \eqref{STPxe} reproduces \eqref{STP}. So  \eqref{STPxe} is equivalent to \eqref{STP}. Notice that   \eqref{STPxe}  is invariant under the reparametrization of the world-line:
\begin{eqnarray}
\tau &\to & \tilde{\tau} = \tilde{\tau}(\tau)\,,\\
e(\tau) &\to & \tilde{e}(\tilde{\tau}) = \frac{d \tau}{d\tilde{\tau}} e(\tau)\,.
\end{eqnarray} 
The reparametrization invariance allows to set
\begin{eqnarray}
e(\tau) &=&  1\,,\\
g_{\mu\nu} \dot{x}^\mu \dot{x}^\nu &=&1 
\end{eqnarray}
as the standard parametrization of the world-line. Doing so yields:
\begin{equation}\label{STPx}
S_{TP}[x] = - m \int d\tau (\frac{1}{2 } g_{\mu\nu} \dot{x}^\mu \dot{x}^\nu + k m  \Phi_\mu \dot{x}^\mu\,),
\end{equation}
We would like to study the theory around slow moving mass distribution.  Consider that the we have a mass distribution composed of $N$ particles. The action of the these particles follow from \eqref{STPx}:
\begin{equation}
S_M \,=\, \sum_{i=1}^N S_{TP}[m_i,x_i] + \sum_{i\neq j} V(i,j) \,,
\end{equation}
where $S_{TP}[m_i,x_i]$ for each particle is given in  \eqref{STPx} and $V(i,j)$ defines the interaction between particles. In the continuum limit, where we have a fluid at equilibrium,  this suggests that 
\begin{eqnarray}
S_M &=& - \int d^4 x \sqrt{-g} (\frac{1}{2}\rho g_{\mu\nu} u^\mu u^\nu + k  \rho \Phi_\nu u^\nu) \\&&+ \text{Interaction of fluid with itself} \,,\nonumber
\end{eqnarray} 
where $\rho$ is the local density of the fluid and $u^\mu$ is the local four velocity of the fluid. Notice that the fluid interaction with itself is not a functional of $\phi_\mu$. In so doing the source current for the vector/gauge field reads:
 \begin{equation}
 J^\mu = k \rho u^\mu. 
 \end{equation}
The equation of motion for the gauge field, in the gauge of $\nabla_\mu \Phi^\mu =0$, reads 
\begin{equation}
\label{MOGPhi}
\Box \Phi^\mu \,=\,- 4 \pi  k J^\mu\,.
\end{equation}
\section{Gravitomagnetism in MOG around the Earth}
In order to address the gravitomagnetism, we look at a small deviation from the flat space-time geometry:
\begin{eqnarray}
g_{\mu\nu} &=& \eta_{\mu\nu} + h_{\mu\nu}\,,\\
\Phi_\mu & =& 0 + \phi_\mu\,, 
\end{eqnarray}
where in the natural unites it holds 
\begin{eqnarray}
 |h_{\mu\nu}|&\ll&1\,,\\
 |\phi_\mu| &\ll&1\,.
\end{eqnarray}
We then consider a slow moving test particle:
\begin{eqnarray}
 |\dot{t}|&\approx&1\,,\\
 |\dot{x}^i| &\ll&c\,.
\end{eqnarray}
These simplify the action of the test particle \eqref{STPx} to 
\begin{equation}
\label{TPEM}
S_{TP} \,=\,m \int dt (\frac{1}{2}  |\dot{x}^i|^2 + \frac{1}{2} h_{00} + k \phi_0 +  (h_{0i} + k \phi^i)\dot{x}^i)\,,
\end{equation}
in analogy with electromagnetism,  the gravitoelectric and gravitomagnetic potentials are then identified to:
\begin{subequations}
\label{EBgravity}
\begin{eqnarray}
\phi_{\text{Phys}} & = & \frac{1}{2} h_{00} + k \phi_0 \,,\\
A^{\text{Phys}}_i & = & h_{0i} + k \phi_i\,.
\end{eqnarray}
\end{subequations}
The equation of motion derived from \eqref{TPEM} can be rewritten as follows
\begin{equation}
\label{appendinxneeds}
\ddot{x} = -\nabla \Phi_{\text{Phys}} +   \frac{v}{c} \times  (\nabla \times A_{\text{Phys}})\,.
\end{equation}
This allows interpreting  $\nabla \times A$ as a gravitomagnetic field. $\nabla \times A$ causes precessions of the orbits of a test particle. This precession is referred to as  the Lense-Thirring precession \cite{Lense}.  Ref. \cite{Iorio:2010rk} provides a decent recent review on  Lense-Thirring precession for planets and satellites in the Solar system. 
The similarity between the gravitomagnetic field and magnetic field beside the spin precession formula in electrodynamics ($\dot{S}= \mu \times B , \mu = \frac{e}{2m} S$)  dictates that  the spin  of a gyroscope  precesses  by  \cite{padi}
\begin{equation}
\Omega_{LT} = - \frac{1}{2}\nabla \times A_{\text{Phys}}\,.
\end{equation}
This precession  is called  the Pugh-Schiff frame-dragging precession \cite{Pugh, Schiff}. The Pugh-Schiff frame-dragging  precession  due to the rotation of the earth recently has been measured by the gravity probe B with the precision of 19\% \cite{Everitt: }.  The gravitomagnetic field due to geodesic effects are measured at an accuracy
of 0.28\%\cite{Everitt: }.  GINGER,  aiming to improve the sensitivity of the ring resonators,   plans  to measure  the gravitomagnetic effect  with a precision at least one order better than that of the gravity probe B \cite{Tartaglia:2012fd}. Also  LAGEOS and LAGEOS 2, and with a number of GRACE (Gravity Recovery and Climate Experiment)  have confirmed the prediction of  General Relativity for the Earth's gravitomagnetic field with with an accuracy of approximately 10\% \cite{LAGEOS}.   Ref. \cite{Murphy:2007nt} shows that   the gravitomagnetic field of the Earth  is in agreement with the Einstein theory's prediction with approximately 0.1\% accuracy via lunar laser ranging (LLR). All the results have been consistent with the Einstein prediction.

We notice that the Einstein's gravitational theory in the Gravitomagnetism approximation holds
\begin{subequations}
\label{EB-EH}
\begin{eqnarray}
\Box \phi_{EH}&=& \frac{1}{2}\Box h_{00}  =  4\pi G \rho \,,\\
\Box A^i_{EH} &=&\Box h_{0i}  =  16 \pi G \rho u^i \,.
\end{eqnarray}
\end{subequations}
While in Moffat's theory of gravity due to  \eqref{EBgravity}, \eqref{EB-EH} and \eqref{MOGPhi} we have:
\begin{subequations}
\label{EB2}
\begin{eqnarray}
\Box \phi_{\text{Phys}}
 & = & 4\pi G_N  \rho \,,\\
\Box A^{\text{Phys}}_i
 & = & 16\pi  G_N  (1 + \frac{3}{4}\alpha) \rho u^i\,.
\end{eqnarray}
\end{subequations}
where \eqref{alphak} are utilised.  The theoretical consistency between Moffat's theory and Einstein theory at the level of gravitomagnetism demands $\alpha=0$. But this converts Moffat's theory to the Einstein theory and renders it useless.  The consistency between Moffat's theory and the results of the Gravity Probe B demands  $|\alpha| < 0.25$ and $|\alpha|<0.0037$ respectively for the measured frame-dragging and the geodetic effects \cite{Everitt: }. Its consistency with LAGEOS and LAGEOS 2, and GRACE (Gravity Recovery and Climate Experiment) demands $|\alpha|<0.13$. Its consistency with the LLR demands $|\alpha|< 0.0013$ \cite{Murphy:2007nt}. 

\section{Conclusion and discussion}
We have reported that the consistency between the MOG gravitomagnetic field and that predicted by the Einstein's gravitional theory and measured by Gravity Probe B, LAGEOS and LAGEOS 2, and with a number of GRACE and Laser Lunar ranging measurements requires  $|\alpha|<0.0013$. 

We notice that there are two estimations for the mass of the supermassive central black hole in M87*: the stellar-dynamical model $(M=6.5 \times 10^9 M_\odot)$ and the gas-dynamical model ($M=3.5 \times 10^9\times M_\odot$). The former mass estimate is consistent with the measured size of the shadow and light emission region of M87* for GR, while the latter estimate is consistent with the MOG prediction with $\alpha=1.13^{0.30}_{-0.24}$ \cite{Moffat:2019uxp}.  We observe that comparing the value of $\alpha$ from the near horizon geometry of M87*'s black hole to the value of $\alpha$ in the solar system is not as simple as it looks. One should first fix $V(G), V(\mu)$ and $V(\Phi)$ in such a way to allow the existence of an interpolating solution from $|\alpha|<0.0013$ in the solar system to its boundary in the Milky Way where $\alpha=O(10)$. The interpolating solution should not contradict any other data at the Solar system as well. Next one should show that the found potentials of the theory allow the existence of an interpolating solution from the boundary of M87* where $\alpha=O(10)$ to the event horizon of its central super massive black hole where $\alpha=1.13^{+0.30}_{-0.24}$.  Taking these steps are outside the scope of the current work and are remained to be addressed. 
\section*{Acknowledgements}
  I would like to thank Atish Dabholkar for the invitation to  ICTP,  ICTP for its nice hospitality, Iva Kordic for converting a large wall in my apartment into a beautiful blackboard. I thank John Moffat for his very valuable feedback and discussion on the paper,  Viktor T. Toth, Niayesh Afshordi, Constantinos Skordis,  Stacy McGaugh, Takeshi Kobayashi, Paolo Creminelli, Loriano Bonora  and Goran Senjanovic for discussions and email correspondences.

 \providecommand{\href}[2]{#2}\begingroup\raggedright
 
\end{document}